\documentclass[prl,aps,showpacs]{revtex4}

\usepackage{amsmath}
\usepackage{graphicx}

\begin{document}

\title{Stretching Instability of Helical Springs }
\author{David A. Kessler}
\affiliation{Dept. of Physics, Bar-Ilan University, Ramat-Gan, Israel}
\author{Yitzhak Rabin}
\affiliation{Dept. of Physics, Bar-Ilan University, Ramat-Gan, Israel}
\date{\today       }

\begin{abstract}
We show that when a gradually increasing tensile force is applied to the
ends of a helical spring with sufficiently large ratios of radius to pitch
and twist to bending rigidity, the end-to-end distance undergoes a sequence
of discontinuous stretching transitions. Subsequent decrease of the force
leads to step-like contraction and hysteresis is observed. For finite
helices, the number of these transitions increases with the number of
helical turns but only one stretching and one contraction instability
survive in the limit of an infinite helix. We calculate the critical line
that separates the region of parameters in which the deformation is
continuous from that in which stretching instabilities occur, and propose
experimental tests of our predictions.
\end{abstract}

\maketitle

The study of mechanical instability of thin elastic rods goes back to Euler
and his contemporaries who considered the buckling of such rods under
compression and torque\cite{Love,TG63}. The present paper presents a
theoretical prediction of a hitherto unknown instability that arises
when a helical spring is stretched by a tensile force applied to its ends.
As we will show in the following, stretching instabilities occur only in
helices whose radius is sufficiently larger than their pitch and whose
rigidity with respect to twist exceeds that with respect to bending.

Consider a helical spring of contour length $L,$ characterized by constant
spontaneous curvature $\kappa _{0}$ and torsion $\tau _{0}$ or,
alternatively, by its radius $r=\kappa _{0}/(\kappa _{0}^{2}+\tau _{0}^{2})$
and pitch $p=2\pi \tau _{0}/(\kappa _{0}^{2}+\tau _{0}^{2})$. Under the
action of a constant stretching force $F$ applied to its ends and directed
along the z-axis (the ends are otherwise unconstrained and no torque is
applied), the helix deforms into a curve whose shape is determined by
minimizing the elastic energy\cite{KT,Nelson,Goldstein,ribbon}%
\begin{equation}
E=\dfrac{1}{2}\sum\limits_{i=1}^{3}a_{i}\int_{0}^{L}ds\left[ \delta \omega
_{i}(s)\right] ^{2}-FR_{z},  \label{energy}
\end{equation}%
with respect to the deviations $\delta \omega _{i}(s)$ of the generalized
curvatures and torsions $\omega _{i}(s)$ from their spontaneous values in
the undeformed state ($s$ is the contour parameter). In general, these
parameters are related to the curvature $\kappa ,$ torsion $\tau ,$ and the
angle $\alpha $ between one of the principal axes of the cross section ($%
\mathbf{t}_{1}$) and the binormal: $\omega _{1}=\kappa \cos \alpha ,$ $%
\omega _{2}=\kappa \sin \alpha $ and $\omega _{3}=\tau +d\alpha /ds.$ We
study here the case that the only contribution to spontaneous twist comes
from torsion ($\alpha _{0}=0)$ and therefore $\omega _{01}=\kappa _{0},$ $%
\omega _{02}=0$ and $\omega _{03}=\tau _{0}$. The coefficients $a_{1}$ and $%
a_{2}$ are bending rigidities associated with the principal axes of inertia
of the (in general, noncircular) cross section and $a_{3}$ is the twist
rigidity. In the following we treat $a_{i}$ as given material parameters of
the spring. The shape of the deformed spring can be obtained by inserting
the $\{\omega _{i}(s)\}$ that minimize the elastic energy into the
generalized Frenet equations for the unit vectors $\mathbf{t}_{i}$ ($\mathbf{%
t}_{3}$ is the tangent to the curve at point $s$ and $\mathbf{t}_{1}$ and $%
\mathbf{t}_{2}$ point along the principal axes of symmetry of the cross
section): 
\begin{equation}
d\mathbf{t}_{i}/ds=-\sum_{jk}e_{ijk}\omega _{j}\mathbf{t}_{k},
\label{Frenet}
\end{equation}%
where $e_{ijk}$ is the antisymmetric unit tensor\cite{Helix}. The space
curve $\mathbf{r}(s)$ associated with the centerline of the deformed spring
is then obtained by integrating the relation $d\mathbf{r/}ds=\mathbf{t}_{3}.$
The system of equations is closed by substituting the expression for the
projection of the end-to-end vector on the z-axis, $R_{z}=\int_{0}^{L}ds%
\mathbf{t}_{3}(s)\cdot \mathbf{z,}$ into Eq. \ref{energy}. Since the spring
will always orient itself along the direction of the force, in the following
we replace $R_{z}\ $by the end-to end-distance $R.$

The force-extension curves are calculated numerically for a helix of 4 turns
by finding a set of $\omega _{i}(s_{j})$ (the continuous curve is replaced
by a discrete set of points, $\left\{ s_{j}\right\} ,$ $j=1,2,...$) that
minimizes the energy, Eq. \ref{energy}. Two typical types of behavior are
found. For sufficiently large ratio of pitch to radius  $p/r=2\pi \tau
_{0}/\kappa _{0}$ and sufficiently small ratio of twist to bending rigidity $%
a_{3}/a_{b}$ (where $a_{b}^{-1}=(a_{1}^{-1}+a_{2}^{-1})/2$), the deformation
is continuous and identical curves are obtained by stretching the undeformed
helix and by starting from a fully stretched configuration and decreasing
the tensile force to zero (Fig. 1). 
\begin{figure}[tbp]
\includegraphics[width=5in]{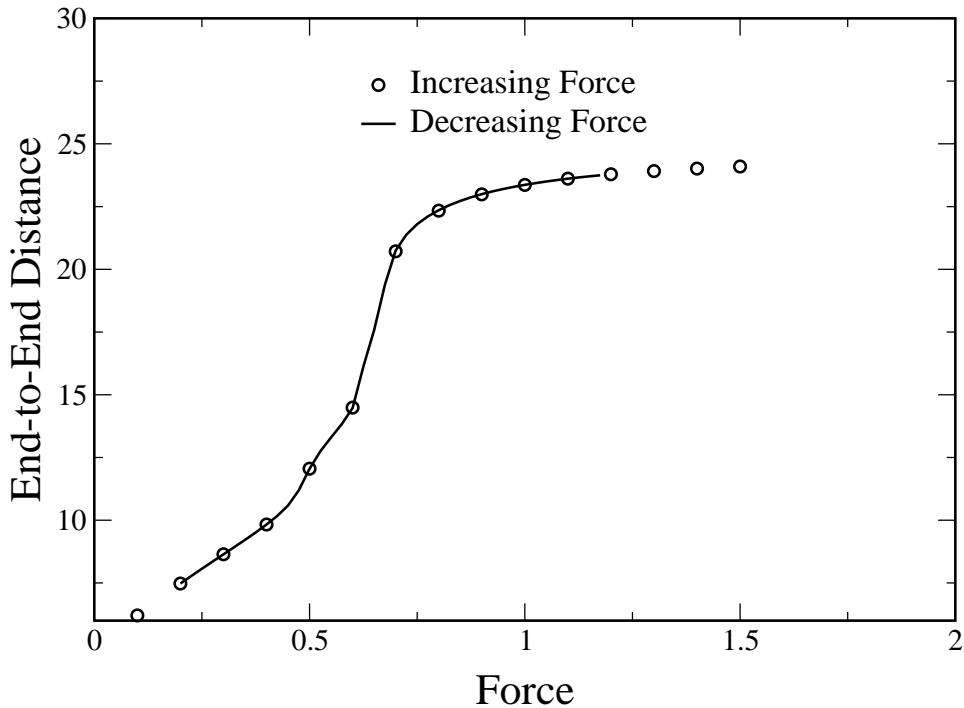}
\caption{Continuous deformation of a helix of $4$ turns with $\protect\kappa %
_{0}=1$, $\protect\tau _{0}=0.2$, $a_{1}=1$, $a_{2}=5$, $a_{3}=3$. We show
the end-to-end distance as a function of the force for both increasing
(circles) and decreasing (solid line) force.}
\end{figure}

In the opposite limit (sufficiently large $r/p$ and $a_{3}/a_{b}$ ratios),
increasing $F$ results in a sequence of discontinuous upward jumps of the
end-to-end distance (Fig. 2), whose number depends on the number of helical
turns and on the increment of the force. Comparison of the conformations on
both sides of a discontinuity shows that the jumps are associated with the
elimination of helical turns (Fig. 2, insets). Discontinuous downward jumps
of the end-to-end distance are also observed when one starts from a fully
stretched configuration and gradually decreases the tensile force but the
locations of these jumps are, in general, different from those of the upward
ones. The observation of hysteresis loops of stretching and contraction
suggests an analogy with first order phase transitions and implies that the
elastic energy $E(R)$ has multiple local minima whose depth and location
vary with the tensile force $F$. In the absence of thermal fluctuations, a
transition to a new energy minimum (not necessarily the lowest energy one)
takes place when the minimum corresponding to the original state disappears.
While in systems undergoing phase transitions the choice of the final state
is determined by the kinetics of the temperature quench, in our case it
depends on the magnitude of the increment of the force.

\begin{figure}[tbp]
\includegraphics[width=5in]{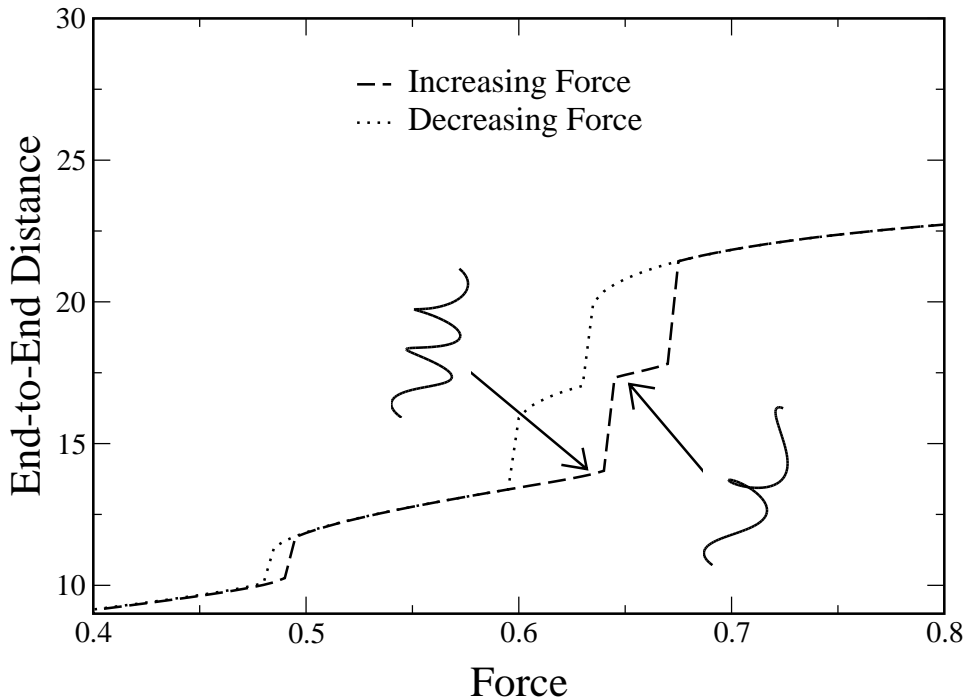}
\caption{Hysteresis loop for a helix of $4$ turns with $\protect\kappa %
_{0}=1 $, $\protect\tau _{0}=0.2$, $a_{1}=1$, $a_{2}=5$, $a_{3}=4$. We show
the end-to-end distance as a function of the force for both increasing and
decreasing force. }
\end{figure}

The analogy with phase transitions suggests that there exists a critical
surface in the parameter space that separates between regions in which the
deformation is continuous from those in which a discontinuous stretching
transition takes place. It is of particular interest to locate this critical
surface in the limit of an infinitely long helix, for which end effects are
negligible and it is expected that the only stable states are perturbations
of the original undeformed helix and of the completely stretched spring (and
therefore only one hysteresis loop that corresponds to a single stretching
and a single contraction instabilities, is expected). Insight into this
limit can be achieved via analytical approximations. In order to reduce the
number of variables, we consider a ribbon with highly asymmetric cross
section, $a_{2}/a_{1}\rightarrow \infty ,$ in which case the cross section
of the ribbon is pinned to the Frenet frame of its centerline, $\alpha (s)=0,
$ and the deformed spring is completely characterized by $\kappa $ and $\tau
.$ Based on our numerical studies of short helices, we would like to stress
that varying $a_{2}$ (while keeping all other parameters fixed) affects
the detailed extension versus force curves but did not change the character
of the transitions, and stretching instabilities accompanied by hysteresis
were observed even in the symmetric case, $a_{1}=a_{2}.$

%In the following we measure the energy in units of $a_{1}\kappa _{0}^{2}L,$
%force in units of $a_{1}\kappa _{0}^{2},$ and distance in units of 
%$1/\kappa_0$.
In the following we measure the energy in units of $a_{1}\kappa _{0}^{2}L$,
and force in units of $a_{1}\kappa _{0}^{2}$. 
In order to obtain analytical results, we consider the limit of an
infinitely long helix and assume that the spring maintains a helical shape
and responds to deformation only by adjusting its curvature and torsion
parameters. Thus, we look for constant ($s$-independent) parameters $\kappa $
and $\tau $ that minimize the energy per unit length, 
\begin{equation}
E/L=\frac{1}{2}\left( \kappa /\kappa _{0}-1\right) ^{2}+\frac{a_{3}/a_{1}}{2}%
\left( \tau /\kappa _{0}-\tau _{0}/\kappa _{0}\right) ^{2}-\frac{F\tau
/\kappa _{0}}{\sqrt{\left( \kappa /\kappa _{0}\right) ^{2}+\left( \tau
/\kappa _{0}\right) ^{2}}}.  \label{Varenergy}
\end{equation}
Support for this variational approximation comes from numerical minimization
of the energy, Eq. \ref{energy}, which shows that while the curvature and
torsion oscillate about their mean values, the amplitude of oscillation
decreases with the length of the spring. In Fig. 3 we compare the extension
vs. force curve for a helix of 12 turns, obtained by exact numerical
minimization of Eq. \ref{energy}, with that obtained using the variational
estimate. The fact that the two curves coincide up to the stability limits
of the initial and the final states provides further support for the
validity of our variational approach.

\begin{figure}[tbp]
\includegraphics[width=5in]{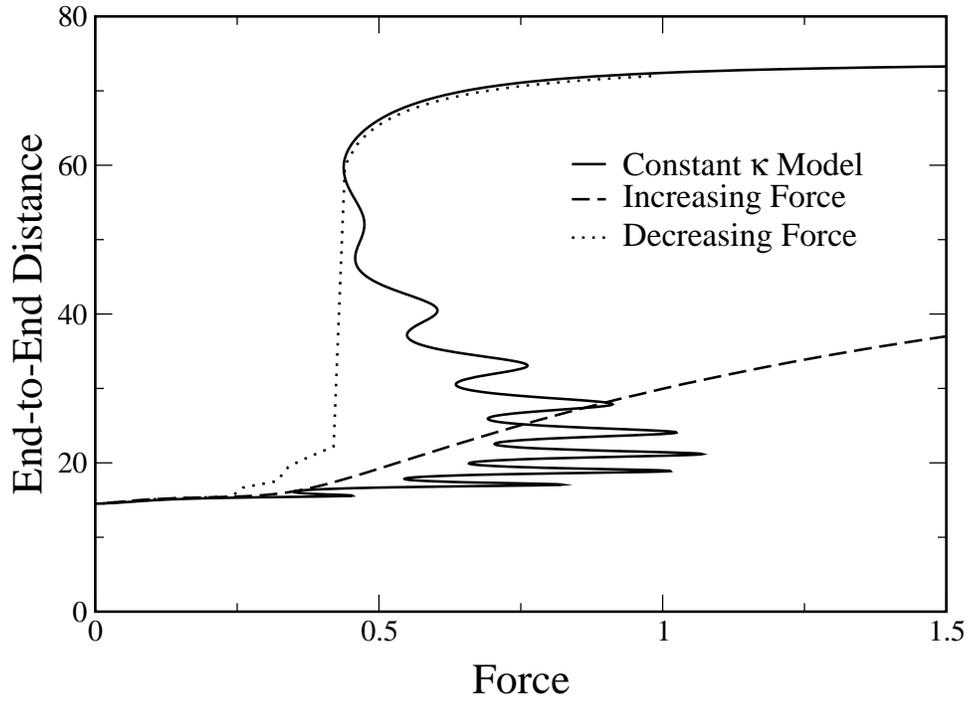}
\caption{Hysteresis loop for a helix of $12$ turns, with $\protect\kappa %
_{0}=1$, $\protect\tau _{0}=0.2$, $a_{1}=1$, $a_{2}=a_{3}=\infty $. We show
the end-to-end distance as a function of the force for both increasing
(dashed line) and decreasing (dotted line) force. Also shown is the result
of our simple model with constant $\protect\kappa $ (solid line).}
\end{figure}

Analytical results can be obtained in several limiting cases. In the case of
large twist rigidity, $a_{3}/a_{1}\rightarrow \infty ,$ the torsion is
pinned to its spontaneous value ($\tau =\tau _{0}$) and the curvature obeys
the relation $a_{1}\left( \kappa /\kappa _{0}-1\right) +F\left( \tau
_{0}/\kappa _{0}\right) \left( \kappa /\kappa _{0}\right) \left[ \left(
\kappa /\kappa _{0}\right) ^{2}+\left( \tau /\kappa _{0}\right) ^{2}\right]
^{-3/2}=0.$ Graphical analysis of this equation shows that it admits either
a single minimum or two minima separated by a maximum. For $\tau _{0}>\tau
_{0,c}$ there is a single minimum of $\kappa $ as a function of $F.$ For $%
\tau _{0}<\tau _{0,c},$ there is a window of $F$ between which there are 3
solutions and outside which there is only one solution. This window closes
at the critical point: $\tau _{0,c}/\kappa _{0}=\left( 2/3\right)
^{5/3}=0.363,$ $F_{c}=0.651,$ and $\kappa _{c}/\kappa _{0}=0.444.$ Another
limit in which the critical point can be calculated analytically for
arbitrary $a_{3}/a_{1}$ is $\tau _{0}/\kappa _{0}\ll 1.$ In this case we
find that two minima exist for $a_{3}/a_{1}>4/3.$ The critical value of this
ratio is therefore $4/3$ and at this point $F_{c}=4/3,$ and $\kappa
_{c}/\kappa _{0}\simeq 54^{1/5}\left( \tau _{0}/\kappa _{0}\right) ^{2/5}/2.$
In between these two limits the critical points have to be found
numerically, resulting in a critical line in the $a_{3}/a_{1}-\tau
_{0}/\kappa _{0}$ plane (Fig. 4). 

\begin{figure}[tbp]
\includegraphics[width=5in]{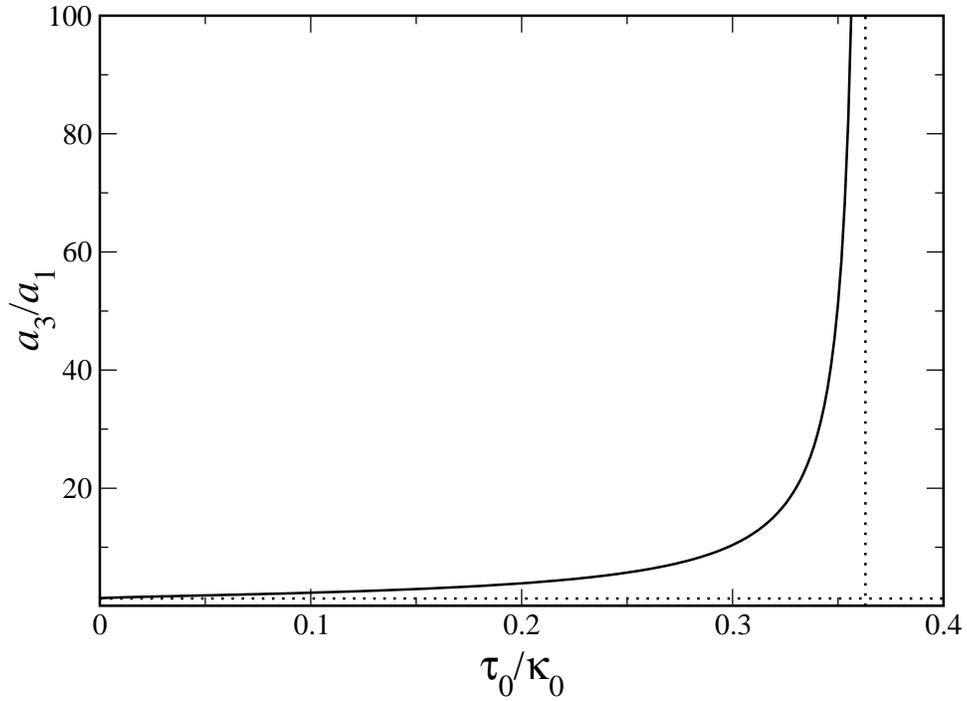}
\caption{ Critical line in the $a_{3}/a_{1}-$ $\protect\tau _{0}/\protect%
\kappa _{0}$ plane for an infinitely long helix with $a_{2}/a_{1}\rightarrow
\infty .$ The vertical and horizontal dotted lines correspond to the two
limits discussed in the text, $\protect\tau _{0,c}/\protect\kappa _{0}=0.363$
and $a_{3}/a_{1}=1.333,$ respectively.}
\end{figure}

We have shown that there exists a range of spontaneous curvatures and
torsions and bending and twist rigidities in which a helical spring does not
deform continuously with the tensile force. Instead, as the force is
increased, the spring undergoes a sequence of stretching instabilities.
Hysteresis is predicted to take place when the force is decreased starting
from full extension, as both the number of sudden contractions and their
locations differ from that of the stretching instabilities at increasing
force. We presented a variational calculation of the critical line that
separates the region of parameters in which the deformation varies
continuously with the force from that in which stretching instabilities take
place. Even though the analysis was restricted to the limit of an infinitely
long spring with a highly asymmetric cross section, our numerical
results indicate that stretching instabilities will be observed in helical
springs with both the radius to pitch ratio and the ratio of twist rigidity
to bending rigidity larger than some numbers of order unity. While the
former requirement can be realized by preparing a spring of a desired
shape, the latter condition may require the use of materials with special
elastic properties. Notice that for isotropic materials the rigidities can
be expressed in terms of the shear and Young moduli $\mu $ and $E$ and the
geometry of the cross section; for example, in the case of rectangular cross
section with sides $c_{1}$ and $c_{2}$ one has\cite{Landau} $%
a_{1}=Ec_{1}c_{2}^{3}/12$, $a_{2}=Ec_{1}^{3}c_{2}/12$, and\ $a_{3}\simeq \mu
c_{1}c_{2}^{3}/3$ (the last expression holds in the limit $c_{1}\gg c_{2}$).
For incompressible materials this yields $a_{3}/a_{1}\simeq $ $4/3$ which
coincides with the critical value of the ratio. The ratio $a_{3}/a_{1}$ can
be increased beyond this critical value by reducing the Poisson ratio or by
using anisotropic materials with high resistance to twist.

We would like to comment on the connection between the stretching
instabilities of helical springs and those familiar from other branches of
physics. The occurrence of multiple energy minima and hysteresis bring to
mind first order phase transitions but since thermal fluctuations and
entropy play no role in our work, the analogy is somewhat misleading. The
predicted transitions appear to be more closely related to hydrodynamic
instabilities, even though the latter arise due to nonlinear effects, while
the former can be described within the framework of the linear theory of
elasticity\cite{comment}. Further reflection shows that the elastic theory
of thin rods is truly linear only in the ``Lagrangian'' frame of the
deformed object (i.e., in its intrinsic coordinate description). In the
presence of external force one must transform into the laboratory frame
(e.g., Euler angles) and the energy is no longer a quadratic form in the
deviations from the undeformed state (recall that hydrodynamics is also a
linear theory in the Lagrangian frame!).

Finally, we would like to stress that since the present analysis is purely
mechanical, the results are directly applicable only to
macroscopic objects. The relevance of the present work to the
deformation of nano-objects such as biopolymers (DNA, proteins), protein
filaments (actin, microtubules, etc.), and carbon nanotubes depends on the
question of whether these objects possess spontaneous curvature and torsion.
Interestingly, discontinuous stretching transitions were reported in single
molecule extension studies of torsionally constrained DNA at high degree of
supercoiling (unlike our model, supercoiling was not spontaneous but was
achieved by the application of torque)\cite{Bensimon}. The study of
stretching instabilities of such objects requires consideration of thermal
fluctuations and extension of the present analysis using the methods of
references \cite{Helix} and \cite{ribbon}, which we hope to report on soon.

\begin{acknowledgments}
DAK and YR acknowledge the support of the Israel Science Foundation.
\end{acknowledgments}


\begin{thebibliography}{99}
\bibitem{comment} Recall that for thin rods linear elasticity applies even
for large global deformations af a long rod, and that the only limitation is
that the characteristic length scales associated with the deformation ($%
1/|\omega _{i}|$) are much larger than the diameter of the spring.
\end{thebibliography}
\end{document}